\documentclass[amsmath,amssymb,aps,prl,superscriptaddress,twocolumn,longbibliography]{revtex4-2}

\usepackage{graphicx}

\usepackage{dcolumn}
\usepackage{bm}
\usepackage{comment}
\usepackage{hyperref}
\usepackage{xcolor}    
\usepackage{float}
\usepackage{mathtools}
\usepackage{multirow}
\usepackage[normalem]{ulem}
\usepackage{relsize}

\newcounter{para}
\newcommand{\para}{\par\refstepcounter{para}\textbf{{\color{cyan}[\thepara]}}\space}

\newcommand{\xEwha}{Department of Physics, Ewha Womans University, Seoul, South Korea}
\let\para\relax

\begin{document}

\title{Materials Expert-Artificial Intelligence for Materials Discovery}

\author{Yanjun Liu}
\affiliation{Department of Physics, Cornell University, Ithaca, NY 14853, USA}
\author{Milena Jovanovic}
\affiliation{Department of Chemistry, Princeton University, Princeton, New Jersey 08544, United States}
\affiliation{Department of Chemistry, North Carolina State University, Raleigh, NC, USA}
\author{Krishnanand Mallayya}
\affiliation{Department of Physics, Cornell University, Ithaca, NY 14853, USA}
\author{Wesley J. Maddox} 
\author{Andrew Gordon Wilson}
\affiliation{Department of Physics, Cornell University, Ithaca, NY 14853, USA}
\affiliation{New York University, New York, NY 10011, USA}

\author{Sebastian Klemenz}
\affiliation{Fraunhofer Research Institution for Materials Recycling and Resource Strategies IWKS, 64357 Hanau, Germany}
\author{Leslie M. Schoop}
\affiliation{Department of Chemistry, Princeton University, Princeton, New Jersey 08544, United States}
\author{Eun-Ah~Kim}
\affiliation{Department of Physics, Cornell University, Ithaca, NY 14853, USA}
\affiliation{\xEwha}

\begin{abstract}
The advent of material databases provides an unprecedented opportunity to uncover predictive descriptors for emergent material properties from vast data space. 
However, common reliance on high-throughput ab initio data necessarily inherits limitations of such data: mismatch with experiments. On the other hand, experimental decisions are often guided by an expert's intuition honed from experiences that are rarely articulated. We propose using machine learning to ``bottle'' such operational intuition into quantifiable descriptors using expertly curated measurement-based data. We introduce ``Materials Expert-Artificial Intelligence'' (ME-AI) to encapsulate and articulate this human intuition. As a first step towards such a program, we focus on the topological semimetal (TSM) among square-net materials as the property inspired by the expert-identified descriptor based on structural information: the tolerance factor. We start by curating a dataset encompassing 12 primary features of 879 square-net materials, using experimental data whenever possible. We then use Dirichlet-based Gaussian process regression using a specialized kernel to reveal composite descriptors for square-net topological semimetals. The ME-AI learned descriptors independently reproduce expert intuition and expand upon it. Specifically, new descriptors point to hypervalency as a critical chemical feature predicting TSM within square-net compounds. Our success with a carefully defined problem points to the “machine bottling human insight” approach as promising for machine learning-aided material discovery. 

\end{abstract}
\maketitle

\para Global efforts of the last decade established vast databases of materials \cite{Horton2021NatComputSci,Landis2012ComputSciEng,Jain2013APLMater,Saal2013JOMa,DeJong2015SciData,Ashton2017PhysRevLett,Haastrup20182DMater,Zhou2019SciData,Draxl2019JPhysMater,Talirz2020SciData,Choudhary2020npjComputMater}  allowing researchers to query data associated with individual entries. This coming age of materials data presents an unprecedented opportunity to discover descriptors of physical properties by inferring and articulating correlations and trends from the collection of entries. If successful, such efforts can lead to guided discovery of new materials. Historically, the most successful of such efforts was that by Mendelev, who predicted properties of elemental materials from the empirical trends articulated in the periodic table. However, discoveries of atomistic features that capture the physical properties of non-elemental materials remain challenging. The combinatorial phase space of non-elemental materials and the large dimensionality of the feature space render the quest out of manual reach.

\begin{figure*}[h]
    \centering
    \includegraphics[width=0.9\textwidth]{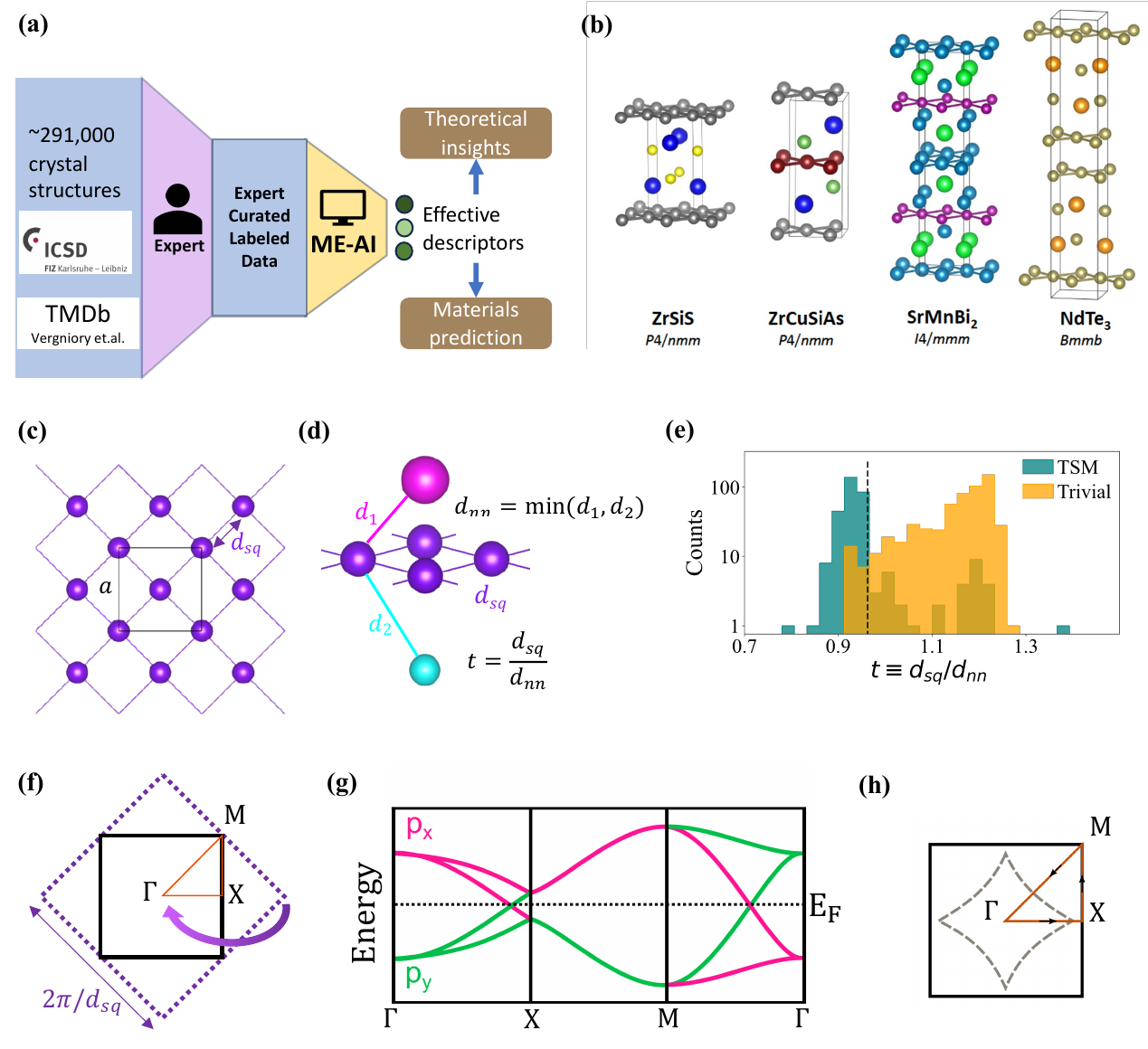}
       \caption{``Materials Expert-Artificial Intelligence'' (ME-AI) for topological semimetals in square-net materials.  {\bf(a):} The conceptual framework of ME-AI.  {\bf(b):} Crystal structure of some 2D square-net materials. 
{\bf(c):} The square-net structure, where purple atoms are arranged on a dense 2D square lattice of lattice constant $d_{sq}$. The centered square-net has an enlarged lattice of lattice constant $a=\sqrt{2}d_{sq}$ containing two atoms. 
{\bf{(d)}:} The 3D square-net structure where the magenta and cyan atoms lie in the out-of-plane directions separated by distances $d_1$ and $d_2$ respectively from the square-net atoms (purple). The tolerance factor  $t\equiv d_{sq}/d_{nn}$ where $d_{nn}\equiv\min(d_1,d_2)$ is the out-of-plane nearest neighbor distance. {\bf(e):} The distribution of tolerance factor ($t$) for the compounds in our data set labeled as TSMs (teal) and trivial (orange) materials, where $t\equiv d_{sq}/d_{nn}$ and $d_{nn}$ is the out-of-plane nearest neighbor distance. We find $t\approx 0.96$ (dashed line) as the value that separates TSMs and trivial materials with maximum accuracy. 
{\bf(f):} The Brillouin zone (BZ) of the enlarged lattice is obtained by folding the larger BZ of the dense lattice. {\bf(g):} The spin degenerate band structure arising from $p_{x}$ (pink) and $p_{y}$ (green) orbitals on the 2D square, folded to the BZ of the enlarged unit cell, and plotted along the path shown in panel (h). Band folding gives rise to two crossings with band inversion (at $\Gamma\rightarrow \text{X}$ and $\text{M}\rightarrow\Gamma$), as well as overlapping bands along the folding edge ($\text{X}\rightarrow\text{M}$). {\bf(h):} BZ with the nodal line (dashed curve) of points where the bands cross. Also shown is the path for the band diagram in panel (g). }
    \label{Fig:TSM}
\end{figure*}

\para With the increasing availability of machine learning tools, recent developments aim at finding stable materials by scaling up ab-initio calculations\cite{Saal2013JOMa,Merchant2023Nature} and synthesis\cite{Stanev2021CommunMater,Gregoire2023NatSynth}. 
However, using machine learning to discover insightful descriptors that connect atoms and structures to desired properties would bypass an exhaustive search. 
There have been fruitful pioneering efforts at learning material descriptors for basic physical and structural properties \cite{Ghiringhelli2015PhysRevLetta,Ouyang2018PhysRevMaterials,Weng2020NatCommuna,Han2021NatCommuna,Damewood2023AnnuRevMaterRes}. 
However, existing efforts have not reached the electronic properties of quantum materials. 
Some electronic properties will be out of reach of atomistic and structural properties. However, an expert human researcher often gleans correlations from empirical experiences to form an intuition that guides their exploration. Hence, it is enticing to ask if we can develop an approach that can ``bottle'' an expert researcher's intuition and go beyond it when supplied with the empirical data that formed the basis of the human researcher's reasoning. Here, we develop such an approach, ME-AI (see Fig.~\ref{Fig:TSM}(a)), and showcase ME-AI's discovery of emergent descriptors for predicting topological materials from the class of compounds with a centered square-net structure.   

\para Given the challenging nature of the problem of material prediction,  the likelihood of success will depend on the reliability of the data and the scope of the search. Since high-throughput $ab$ $initio$ simulation data tend to lack accuracy, foundational data should come from measurements. As for the scope of the search, setting the scope that makes ample use of chemical understanding, i.e., using a strong prior,  can increase the likelihood of success. Based on these two criteria, we focus on learning the descriptor of topological nodal-line semimetals derived from the basic atomistic features of the elements and structural features, limiting the pool of candidate materials to square-net semimetals.  

\para
Topological nodal-line semimetals  (TSMs) are characterized by their band structures, in which two spin-degenerate bands cross along a line in momentum space. They are studied for their novel physical properties and possible application in energy conversion \cite{Fu2018EnergyEnvironSci}, as electrocatalysts \cite{Xie2022SmallScience}, and for sensors \cite{Liu2020NatMater}. Conventional identification of TSMs requires a detailed analysis of materials' symmetry through the derivation of band representations and the number of electron filling \cite{Bradlyn2017Natureb}. 
However, some of us observed that chemical intuition can be instructive in predicting TSMs among 2D ``square-net''  materials.  
The ``square-net''  materials are crystalline solids that have the structural motif of a 2D centered square net, also known as a $4^4$ net (Fig.~\ref{Fig:TSM}(b,c)). 
Following chemist's intuition, some of us introduced a structural descriptor called ``tolerance factor ($t$-factor)'' in Ref. \cite{Klemenz2019AnnuRevMaterResa, Klemenz2020JAmChemSoca}, as the ratio of square lattice distance to the out-of-plane nearest neighbor distance (Fig. \ref{Fig:TSM}(d)). The $t$-factor does a remarkable job at distinguishing TSMs, which have smaller values of $t$, from trivial materials  (see Fig. \ref{Fig:TSM}(e)). However,  at intermediate values of $t\sim1$, it is clear that additional descriptors are necessary. The simplicity of the  $t$-factor and strong domain knowledge incorporated into the choice of  2D square-net compounds 
 make this problem an ideal starting point for developing and testing ME-AI.

 \begin{figure*}[ht]
    \centering
    \includegraphics[width=0.9\textwidth]{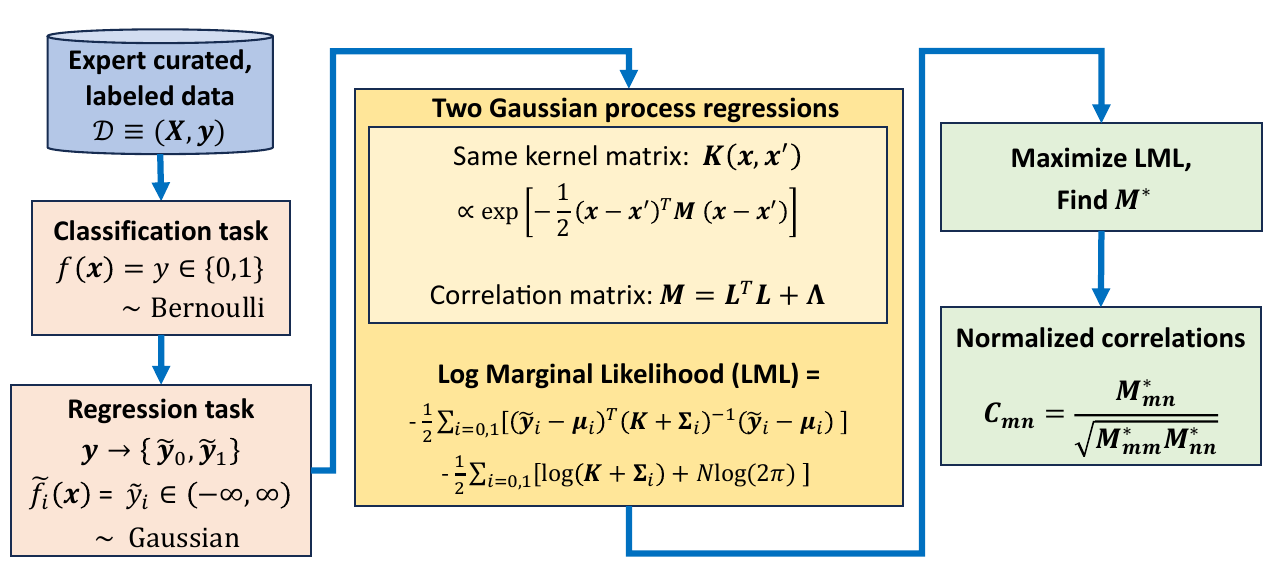}
    \caption{ A flowchart illustration of ME-AI. We first create a preprocessed and labeled data set $\mathcal{D}\equiv (\mathbf{X},\mathbf{y})$ composed of the 12 primary features, $\mathbf{X}=(\mathbf{x}_1,\dots, \mathbf{x}_N)$ with $\mathbf{x}_n\in R^{D}$ where $D=12$, and their class labels $\mathbf{y}=(y_1,\dots,y_N)$ with $y_n\in \{0,1\}$. The subscript $n$ denotes different materials in the total $N$ entries database. After converting the classification task into two independent regression tasks, we set up two GPRs with the shared kernel $\mathbf{K}(\mathbf{x},\mathbf{x}\,')$. The correlation matrix $\mathbf{M}$ inside the kernel couples the different primary features and is designed to have a factor analysis structure $\mathbf{M}=\mathbf{L}^T\mathbf{L}+\mathbf{\Lambda}$, which allows us to learn interaction between different PF's with a limited number of hyperparameters. We train the model with the whole data set by maximizing the log marginal likelihood. This automatically includes regularization. We get the correlations $\mathbf{M}^*$ from the trained model parameters between the primary features. The significance of the coupling between the $m^{\text{th}}$ and $n^{\text{th}}$ primary features is measured by the normalized correlation $\mathbf{C}_{mn}$. This reveals the pairs of significant features that identify the square-net TSMs. }
    \label{Fig:flowchart}
\end{figure*}

\para For completeness, we first review the structural and chemical insight that led to the $t$-factor. 
 When the 2D square net is considered in isolation, while the atoms are arranged in a lattice of length $d_{sq}$, the contributions from the out-of-plane atoms lead to the centered square-net with an enlarged unit cell of length $\sqrt{2}d_{sq}$ containing two atoms (Fig. \ref{Fig:TSM}(c)). If the perturbative effects of the out-of-plane layers are negligible, the effective band structure can be obtained by folding the tight-binding bands of the square lattice into the smaller Brillouin zone (BZ) of the enlarged lattice (Fig. \ref{Fig:TSM}(f)). The band folding creates crossings, which the symmetry of the enlarged lattice may protect. With the right electron filling to align the Fermi level at the band crossings, the material becomes a TSM. Ref \cite{Klemenz2020PhysRevB} illustrates these points through a minimal tight-binding model on a square-net lattice with 6 electrons per atom. 
With filled $s$ and $p_z$ orbitals and half-filled $p_x$ and $p_y$ orbitals, the folded band structure (Fig. \ref{Fig:TSM}(g)) shows bands crossing at points that lie on a closed continuous curve inside the BZ (Fig. \ref{Fig:TSM}(h)). Having 6 electrons per atom ensures that the Fermi level passes through the band crossings, making this material a TSM. This inspired Klemenz et al. \cite{Klemenz2020JAmChemSoca} to propose a ratio between two structured parameters quantifying deviation from 2D square-net plane structure (see Fig.~\ref{Fig:TSM}(d)), $t{-\rm factor}\equiv d_{sq}/d_{nn}$, where  $d_{nn}$. The distribution of $t$-factor within the database shown in  Fig.~\ref{Fig:TSM}(e) shows that while it is effective at indicating TSM, $t$-factor leaves ambiguity around $t\approx1$.

\para 
ME-AI aims to reveal the correlations between different primary features (PF)s and discover the emergent descriptors composed of primary features. We choose the PFs to be atomistic or structural such that we can interpret the ML results from a chemical perspective. 
For atomistic features, we focus on 
(a) electron affinity, the change in energy after a neutral atom gains an electron; (b) electronegativity, the relative ability of an atom to attract a shared electron pair; and (c) valence electron count, corresponding to the number of electrons in a neutral atom. Since each compound consists of different numbers of elements, we built primary atomistic features in a uniform structure by choosing maximum and minimum values of these features and the square-net elements features (see Table S1). The last atomistic feature is (d) the estimated face-centered cubic lattice parameter $fcc$ of the square-net element, which reflects the radius of the square-net atom.
Given the prior choice of square-net compounds, two lengths are sufficient to represent the structural feature variations: 
(e) the crystallographic characteristic distances $d_{sq}$ and $d_{nn}$.

\para Having determined the 12-dimensional PFs, we curated an experimentally measured PF database of PFs for 879 compounds belonging to the 2D-centered square-net class in the inorganic crystal structure database (ICSD)\cite{LevinICSD}. The structure types of these compounds include PbFCl, ZrSiS (ZrSiTe, AmTe$_{2-x}$, and UP$_2$), PrOI, Cu$_2$Sb, ZrCuSiAs-HfCuSi$_2$, LaZn$_{0.5}$Sb$_2$, PrNiSb$_2$ and CaBe$_2$Ge$_2$.
The expert labeling of the materials relied on the visual comparison of the measured or calculated\cite{Bradlyn2017Natureb,Vergniory2019Natureb} 
band structure to the square-net tight-binding model band structure  Fig.~\ref{Fig:TSM}(g).

\para From the machine learning perspective, the task at hand is to use this relatively small labeled dataset of 12-dimensional PFs to discover effective descriptors that predict TSMs automatically. A conventional approach of feature dimension reduction using principal component analysis (PCA) fails (see SM section B) since PCA cannot incorporate prior knowledge of the labels. On the other hand, the go-to supervised learning architectures using neural networks are also inadequate due to the small dataset, lack of interpretability, and the danger of overfitting.
ME-AI treats this problem as a Gaussian Process (GP) classification task \cite{MiliosNEURIPS2018} and obtains the descriptors by learning a purposefully chosen factor analysis matrix parameterizing the kernel for the GP \cite{VivarelliNIPS1998}.

\para ME-AI starts with an expert-curated database that carries and informs the expert's intuition (see Fig.~\ref{Fig:TSM}(a)). Of the 879 materials, we select 861 materials with structure types other than CaBe$_2$Ge$_2$ as our training data. We exclude the CaBe$_2$Ge$_2$ structure type for training as these entries have errors in their labels (See SM section A for details). To ensure a fair comparison of the PFs with different magnitudes and units, each PF is linearly normalized to lie on the range $[0,1]$ (min-max normalization). Thus, our preprocessed training data set $\mathcal{D}=\left(\mathbf{X},\mathbf{y}\right)$ for the $N=861$ materials consists of the input features (normalized PFs) $\mathbf{X}=\{\mathbf{x}_1,\dots,\mathbf{x}_N\}$ and the output labels  $\mathbf{y}=\bigl(y_1,\dots,y_N\bigr)^{T}$. The features   $\mathbf{x}_n\in\mathbb{R}^D$ is a $D=12$ dimensional  vector of the $n^{\text{th}}$ material, whose output label $y_n\in \{0,1\}$ represents the trivial class as $0$ and TSM class as $1$. The ME-AI proceeded in three steps, as depicted in Fig.~\ref{Fig:flowchart} and SM section C.

{\it{Step 1}}: Assuming a Bernoulli model for the discrete class labels,  we start with a Beta prior for the distribution of labels. Next, we transform these labels to a
latent space where we can approximate the likelihood by a Gaussian, following the route of Ref~\cite{MiliosNEURIPS2018} (see SM section C). With this transformation, each label $y_n\in \{0,1\}$  gives two new variables $\tilde{y}_0, \tilde{y}_1 \in (-\infty,\infty)$ (given by Eq.(15) in SM section C), whose distributions are approximately Gaussian. We can now proceed with two Gaussian Process Regressions (GPRs) for the $\tilde{\mathbf{y}}_0$ and $\tilde{\mathbf{y}}_1$.  

{\it{Step 2:}} As our goal is to learn the feature correlations that encompass both labels, we design a Mahalanabois kernel $\mathbf{K}$ to be shared between the two GPRs, as   \begin{equation}
    \mathbf{K}(\mathbf{x}, \mathbf{x'}):=\sigma_p^2 \exp[-\frac{1}{2}(\mathbf{x}-\mathbf{x'})^T\mathbf{M}(\mathbf{x}-\mathbf{x'})]\mathrm{,}
\end{equation}
whose elements connect the features $\mathbf{x}$ and $\mathbf{x'}$.   Assuming the existence of a smaller number of hidden features emerging from combinations of the primary features \cite{VivarelliNIPS1998}, we use a factor analysis model for the structure of the correlation matrix $\mathbf{M}$ as,  
\begin{eqnarray}
\mathbf{M}=\mathbf{L}^T\mathbf{L}+\mathbf{\Lambda},
\end{eqnarray}
where $\mathbf{L}$ is a $q\times D$ matrix with $q<D$, and $\mathbf{\Lambda}$ is a  positive definite diagonal matrix. The $q$ sets the number of common factors, and we find $q=6$ to be the optimal choice, as found through hyperparameter tuning (see SM section D).   The smaller value of $q$ limits the number of hyperparameters to be learned from our relatively small volume of the data.

{\it Step 3:} Now we carry out the standard GPR by maximizing the log marginal likelihood (LML) to learn the hyperparameters, including the correlation matrix $\mathbf{M}$, which will reveal the descriptors. Given the assumed Gaussian distribution of the transformed variables $\Tilde{\mathbf{y}}_i$ the LML has the usual closed-form expression~\cite{Rasmussen2006a} given by, 
\begin{flalign}
    \text{LML}=&\frac{1}{2}\sum_{i=0}^1\;\Bigl[(\Tilde{\mathbf{y}}_i-\bm{\mu}_i)^T\bigl(\mathbf{K}+\mathbf{\Sigma}_i\bigr)^{-1}(\Tilde{\mathbf{y}}_i-\bm{\mu}_i)\nonumber\\
    &+\mathrm{log}|\mathbf{K}+\mathbf{\Sigma}_i|+N\mathrm{log}2\pi\Bigr],&
\end{flalign}
where $\mathbf{K}$ and $\bm{\mu}_i$ are the parameters to be learned, and $\mathbf{\Sigma}_i$ is a diagonal matrix set by the Beta prior (see SM section C). In the above expression, the first term promotes the model fitness while the second term 
penalizes model complexity. 
The marginal likelihood produces principled and automatic regularization, satisfying the ``Occam's razor'' property \cite{mackay2003information}.  
With our database, training to maximize LML takes $\sim8$ minutes using a Quadro GV100 GPU (see SM section D for details). We interpret the LML maximizing matrix elements  $\mathbf{M}^*$ 
to reflect how PFs interact to form the descriptors of TSM. For this, we construct a normalized correlation with elements given by
\begin{eqnarray}
\mathbf{C}_{mn}=\dfrac{\mathbf{M}^*_{mn}}{\sqrt{\mathbf{M}^*_{mm}\mathbf{M}^*_{nn}}}.
\end{eqnarray}
Now, the correlation matrix reveals the interaction between PFs in forming descriptors of TSM.

\begin{figure*}[htbp]
    \centering
\includegraphics[width=0.92\textwidth]{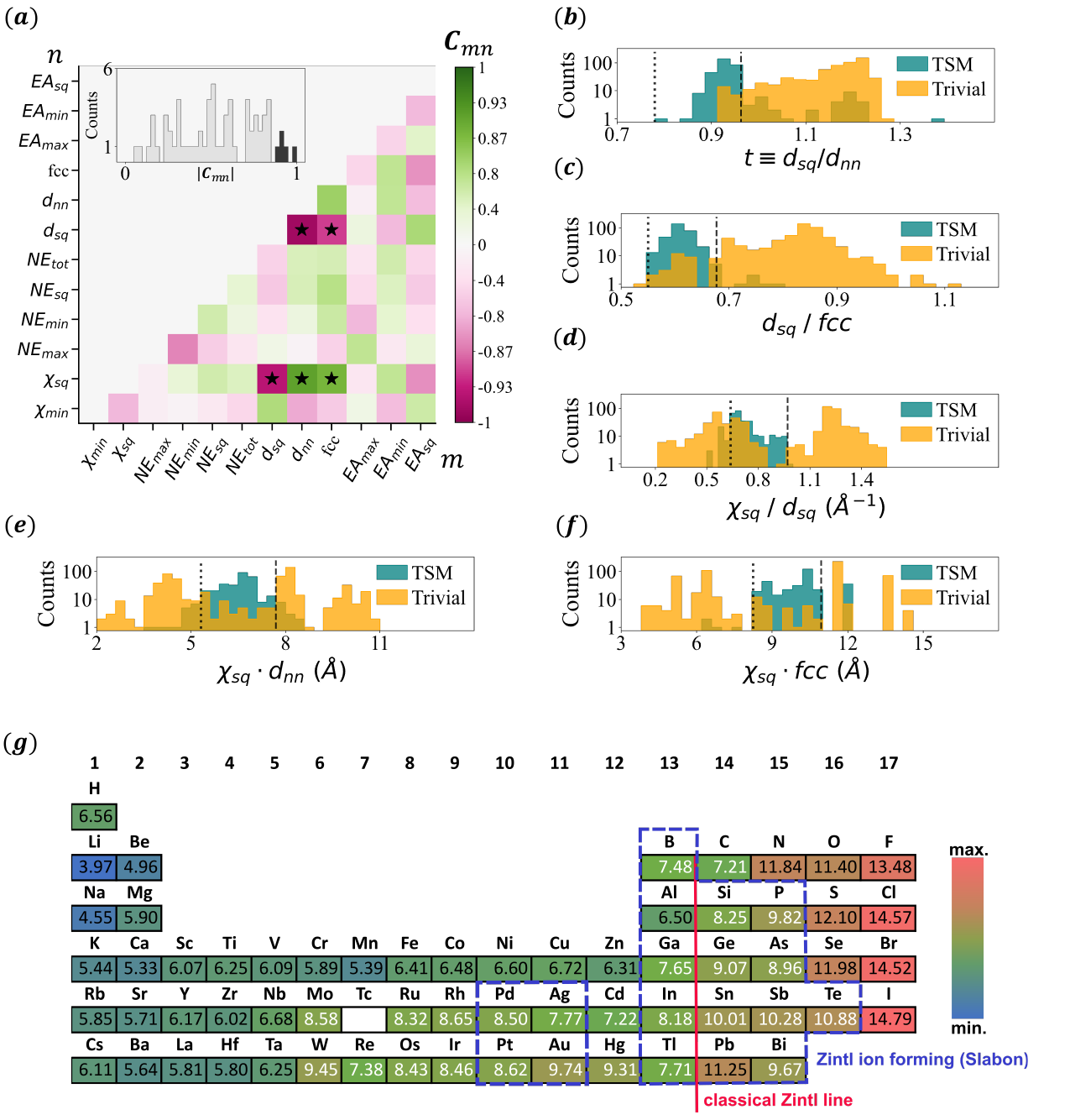}
    \caption{The main results from the GP model. {\bf{(a):}} The normalized-correlation matrix $\boldsymbol{C}_{mn}$, where $m$ and $n$ denote the primary features. Only the lower triangle is shown. The inset shows the distribution of magnitudes of non-diagonal elements, $|\boldsymbol{C}_{mn}|$, with the strongest elements shown in black. These strongest elements: ($d_{sq}$, $d_{nn}$), ($d_{sq}$, $fcc$), ($\chi_{sq}$, $d_{sq}$), ($\chi_{sq}$, $d_{nn}$), and
 ($\chi_{sq}$, $fcc$) are marked with a star. The values of  {\bf{(b):}} $d_{sq}/d_{nn}$, {\bf{(c):}}  $d_{sq}/fcc$, {\bf{(d):}} $\chi_{sq}/d_{nn}$, {\bf{(e):}} $\chi_{sq}\cdot d_{nn}$, and {\bf{(f):}} $\chi_{sq}\cdot fcc$, for the TSM and trivial materials in our dataset. The vertical lines separate TSMs from trivial materials.   {\bf{(g):}}  The periodic table with the new element-specific descriptor $\chi_{sq}\cdot fcc$ color-coded from lowest to highest values from blue to red, respectively. The entries in white font fall in the window of 7 to 11. The red line represents the classical Zintl line, while the blue framed elements are known to form Zintl ions as described by Slabon\cite{slabon2013structure}}
    \label{Fig:results}
\end{figure*}

\para The elements of the normalized correlation matrix, $\{\mathbf{C}_{mn}\}$ shown in Fig.\ref{Fig:results}(a) 
reveal the relative strengths and signs of interaction among all the 72 pairs of PFs. Five elements of $\{\mathbf{C}_{mn}\}$ clearly stand out in their magnitudes with varying signs. 
We hypothesize that positive (negative) and strong $\{\mathbf{C}_{mn}\}$ can be effectively captured by emergent descriptors formed from products (ratios) or PF $m$ and PF $n$. To test this hypothesis, we plot the distribution of the products and ratios,  $\{\mathbf{C}_{mn}\}$, i.e., $d_{sq}/d_{nn}$, $d_{sq}/fcc$, 
 $\chi_{sq}/d_{sq}$, $\chi_{sq}*d_{nn}$, and $\chi_{sq}*fcc$ as shown in Fig.\ref{Fig:results}(b-f). As shown in SM E, the distributions confirm small values of  $d_{sq}/d_{nn}$, $d_{sq}/fcc$ and intermediate values of $\chi_{sq}/d_{sq}$, $\chi_{sq}*fcc$, and $\chi_{sq}*d_{nn}$ as successful descriptors of TSM. These descriptors synthesize atomistic information of elements forming the material and the structural information of the compound in three different ways: 
(1) $d_{sq}/d_{nn}$ relies solely on structural features, (2) $d_{sq}/fcc$, $\chi_{sq}*d_{sq}$, and $\chi_{sq}*d_{nn}$ combine structural and atomistic features, and (3) $\chi_{sq}*fcc$ is a descriptor solely based on atomistic features. It is noteworthy that ME-AI reproduces $t=d_{sq}/d_{nn}$ as one of the emergent descriptors without prior knowledge. This bolsters the insight underlying the $t$-factor and establishes ME-AI's capability to ``bottle'' expert insight latent to the expert curation of the data. 

\para We now turn to four new descriptors.
Among these, the distribution of $d_{sq}/fcc$ (Fig.~\ref{Fig:results}(c))  is similar to the distribution of the $t$-factor (Fig.~\ref{Fig:results}(b)) although  $d_{sq}/fcc$ does a better job in separating TSM from trivial materials (see SM E for quantitative analysis of the distributions). Since $fcc$ estimates the size of the unbonded square net atoms, small values of $d_{sq}/fcc$  and that of the $t$-factor both point to chemically and electronically significant role of the square-net planes. 

\para The distributions of the remaining three descriptors, $\chi_{sq}*d_{nn}$, $\chi_{sq}/d_{sq}$, and $\chi_{sq}*fcc$, are similar in that TSM favors intermediate values of these descriptors (Fig.~\ref{Fig:results}(d-f)). This result reveals the underappreciated significance of hypervalency for TSM. 
The concept of hypervalency accounts for the seeming violation of the octet rule in molecules and solids,  a classic chemical rule 
the main group elements want to reach noble gas configuration with 8 valence electrons to fill the $s$ and $p$ shells~\cite{tro2013chemistry}. 
Hypervalent bonds connect more than two atoms and allow cases with the seeming violation of the octet rule to satisfy the octet rule in an extended sense~\cite{APapoian2000AngewChemIntEd,Rogachev2013InorgChem}. 
While hypervalent bonds were shown to be possible in square-net materials, a connection between hypervalency and TSM had not been made previously. The three descriptors address complimentary but distinct aspects determining hypervalent bonds in square-net atoms. $\chi_{sq}*d_{nn}$ and $\chi_{sq}/d_{sq}$ reveals the bonding tendency with out-of-plane atoms, and $\chi_{sq}*fcc$ reflects the atomistic features that facilitate hypervalent bonding (see SM section F). Specifically, only elements of intermediate electronegativity and size will be inclined to form electron-sharing hypervalent bonds. 

\para Since $\chi_{sq}*fcc$ is made of purely atomistic features of square-net atoms, we can calculate the descriptor for each element of the periodic table, as shown in Fig.~\ref{Fig:results}(g)\footnote{We exclude radioactive elements and rare earth elements.}. The positions of elements in the periodic table with $7<\chi_{sq}*fcc<11$, the range identified to support TSM in Fig.~\ref{Fig:results} (f), reveals a remarkable connection between this ME-AI learned descriptor and the classic chemical concept of ``Zintl line'' \cite{zintl1939intermetallische}. The Zintl-line divides elements that form covalent bonds with each other to yield polyanions (or polycations) from elements that form regular monoatomic cations (or anions) \cite{zintl1939intermetallische, janka2011zintl}. Thus it can also point to hypervelent bonds, as they can be viewed as a type of covalent bond. Classically, the Zintl line is drawn between group 13 and 14 \cite{Wang2011InorgChem}; however, extensions of the Zintl-Klemm concept included the higher homologues of group 13 and heavy transition metals such as Au \cite{janka2011zintl}.  
Thus, the ME-AI learned atomistic descriptor $\chi_{sq}*fcc$ agrees with the extended Zintl-Klemm concept and highlights the importance of covalent bonding for TSM formation.

\begin{figure*}[ht]
    \centering
    \includegraphics[width=0.95\textwidth]{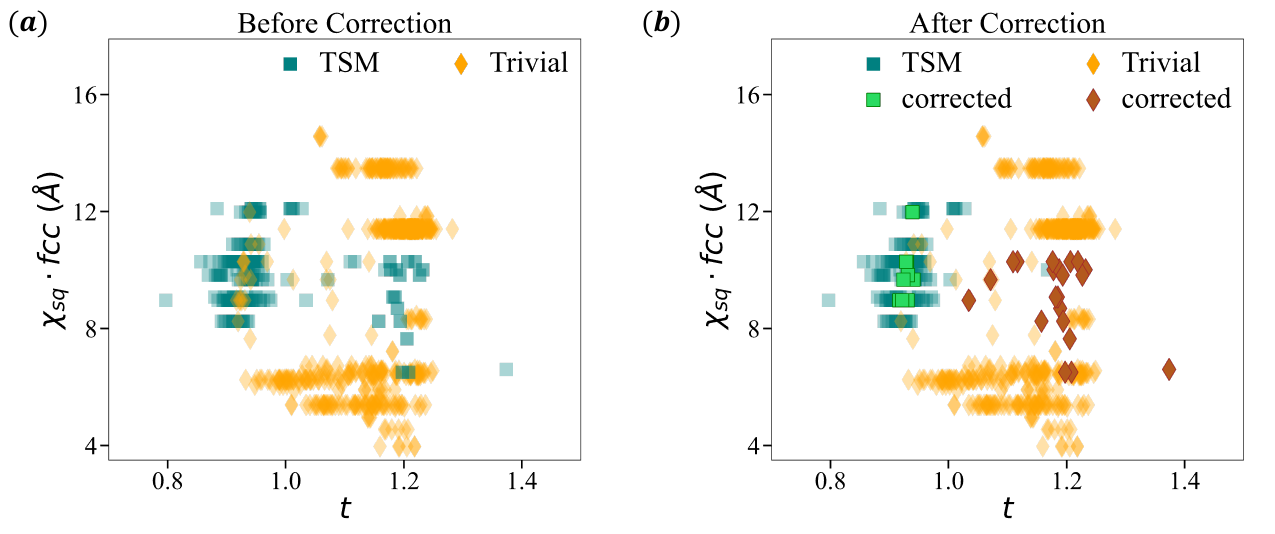}
    \caption{The scatter plot showing TSM and trivial materials in the 2D descriptor space spanned by the $t$-factor and $\chi_{sq}\cdot fcc$. The TSM (teal) and trivial (orange) materials are separated sharply in this 2D space. {\bf{(a):}} The originally labeled TSM (teal) and trivial (orange) materials show some outliers in the clustered regions of TSM and trivial materials. {\bf{(b):}}  Further analysis enabled us to correct mislabeled materials in our original data set. The dark brown points denote the trivial materials originally classified as TSMs; the bright green point denotes a TSM originally classified as trivial.}
    \label{Fig.4}
\end{figure*}

\para Finally, the excellent separation between TSM and trivial materials we achieved with the ME-AI learned descriptors allowed us to detect misplaced labels in the original curation of the material dataset, as shown in Fig.\ref{Fig.4}. With the $t$-factor alone, significant mixing between TSM and Trivial (see Fig.\ref{Fig:TSM}(e)) made it difficult to identify true outliers and potential mislabeling. However, using a two-dimensional descriptor space $(t,\chi_{sq}*fcc)$ as shown in \ref{Fig.4}(a), the TSM and trivial materials separate sharply and draw attention to the 44 outliers. Upon inspection, we found the original labeling left some ambiguity for these outlier compounds. For a closer inspection, we performed in-house DFT calculations on these outliers and compared the orbital-projected band structure with the tight-binding model band structure in Fig.\ref{Fig:TSM}(g). For 48 compounds, this close-up study resulted in label correction. The corrected distribution in \ref{Fig.4}(b) demonstrates the power of the ME-AI approach in bottling the essence of the expert intuition in a robust manner that can withstand small unintended errors inevitable in data curation. 

\para In summary, we introduced a machine learning strategy, ME-AI, that bottles expert intuition underlying expertly curated data and articulates the intuition into quantifiable descriptors for target features. The strategy starts with expertly curated and labeled data. Deployed to a 12-dimensional experimentally measured PF set for an expertly curated and labeled dataset of 879 square-net compounds, ME-AI rediscovered $t$-factor that human experts gleaned from the same data independently. Moreover, ME-AI revealed four new emergent descriptors. 
While one of the new descriptors addresses the same chemical intuition as the $t$-factor, highlighting the critical role of square-net plane, the remaining three new descriptors all point to the new insight regarding hypervalency. 
In particular, one of the new descriptors $\chi_{sq}*fcc$ is entirely atomistic concerning the square-net atoms, allowing us to place the new descriptor on the periodic table and check the hypervalency-interpretation against chemical knowledge. The success demonstrates ME-AI's capability to encapsulate human chemical intuition and expertise from relatively small, expertly curated data to advance materials discovery. Such synergy between human experts and machine intelligence can accelerate materials discovery, building on precious expert experiences.


\vspace{5mm}

{\bf Acknowledgement.} MJ, WM, AW, LS, and EAK were supported by NSF grant OAC-2118310. 
The computation was done using high-powered computing
cluster that was established through the support of the Gordon and Betty Moore Foundation’s EPiQS Initiative, Grant GBMF10436 to EAK, and through the support of the MURI grant FA9550-21-1-0429. YL was supported by the Gordon and Betty Moore Foundation’s EPiQS Initiative, Grant GBMF10436 to EAK, and by the  MURI grant FA9550-21-1-0429. KM was supported by Eric and Wendy Schmidt AI in Science Postdoctoral Fellowship: a Schmidt Futures program.
S.K. was supported by the Fraunhofer Internal Programs under Grant No. Attract 170-600006.

\clearpage


\newpage
\bibliographystyle{apsrev4-2}
\bibliography{TSMRefs}

\clearpage

\pagebreak
\clearpage
\appendix
\onecolumngrid
\renewcommand{\thefigure}{S\arabic{figure}}
\renewcommand{\thetable}{S\arabic{table}}

\section{Supplementary Material}

\section{SM A: Initial Data handling}\label{app:data}
\begin{table}[ht]
 \centering
    
  \includegraphics[width=0.7\textwidth]{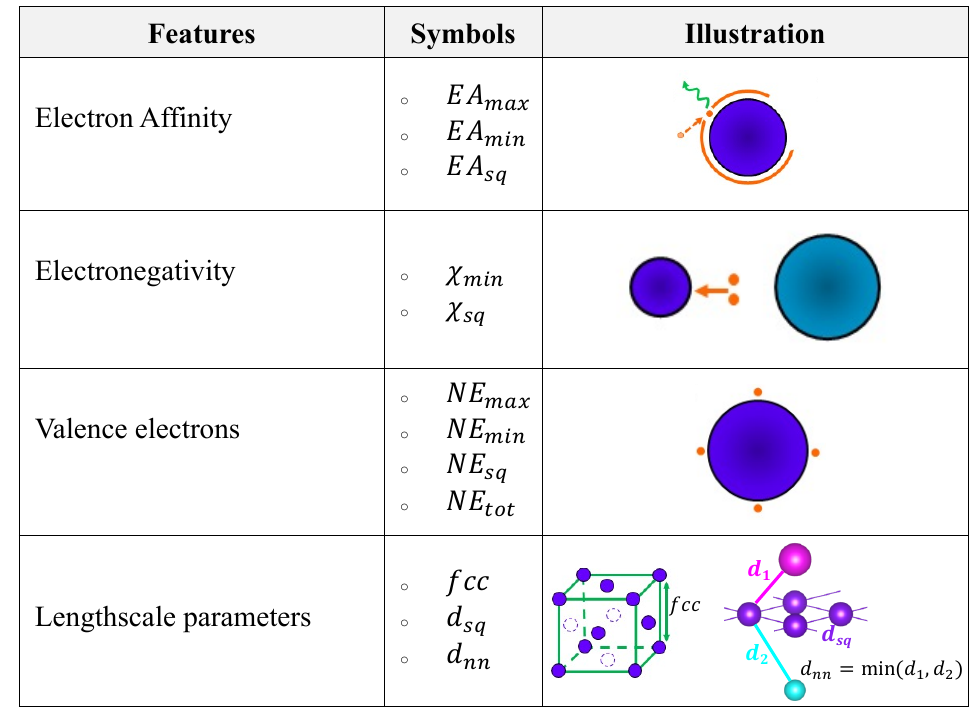}
  \caption{Primary features. Subscripts $max$, $min$, and $sq$ refer to the maximum, minimum, and square-net element values. The subscript $tot$ refers to the total number of electrons based on the stoichiometric information. We omit $\chi_{max}$ because of its high covariance with $\chi_{sq}$. Graphical illustrations of the atomic features are shown in the third column.}.
  
  \label{tab:features}
    \end{table}

The initial data set is based on the data set curated for \cite{Klemenz2020JAmChemSoca}. The data presented in this work was compiled from structural information in the Inorganic Crystal Structure Database (ICSD) \cite{LevinICSD} and Topological Materials Database (TMDb) \cite{Vergniory2019Natureb}. The initial data set was compiled using these sources for structural and topological information, respectively. Data from the TMDb was collected through March 2019 and from the ICSD through December 2019. Only those structure types in space group 129 that exhibit a $4^4$-net were considered. The ICSD entries were sorted by structure type and double entries were removed. If two entries described the same compound (unique chemical composition), the entry with the measurement conditions closest to standard conditions or with the most accuracy was chosen. For structures that contain more than one square net, where both are composed of main group elements, only the lowest tolerance factor $t$ is given.

Only the entries in space group 129 were expanded on for this work. In the first step the entries were compared to band structure data from TMDb \cite{Vergniory2019Natureb}. The presence of the iconic band structure feature seen in 4\textsuperscript{4}-net compounds were identified visually.
The visual identification relied on the manual comparision 
between the DFT band structures in the TMDb \cite{Vergniory2019Natureb} and 
the tight-binding band structure shown in Fig 1 (g).
If bands that visually looked like the tight-binding bands were found in the DFT band structure around the Fermi level, the compound was labeled as TSM. 

The compounds were categorized as 'yes' if they had clean and obvious square-net TSM bands, and as 'no' if they obviously lacked TSM bands. If the compounds had complicated but likely square-net TSM bands, or if their band structures were not reported, but similar compounds are known to be TSMs, they were categorized as "yes*" but with less confidence. On the other hand, if the compounds had complicated but likely topological trivial bands, or if their band structures were not reported, but similar compounds are trivial materials, they were categorized as "no*". Compounds with mixed-occupation or vacancies that did not have an entry in the TMDb were categorized according to the entry with the same structure type and closest composition. To each compound, additional data about the elements included was compiled, such as electron affinity ($EA$) and electronegativity ($\chi$) from CRC Handbook of Chemistry and Physics, estimated face-centered cubic lattice parameter ($fcc$) from the Xenopy database and valence electron count ($NE$). 

Because the materials in our data set have various structures and compositions, we need to determine how we represent them with the atomic features. The basic idea is to use the maximum and minimum values among each material's elements. Here, we also include values of square-net atoms to account for this specific structure character. Before we group these features into the feature matrix representation, we eliminate $\chi_{max}$ because of its high covariance with $\chi_{sq}$. We consider only the square-net atoms' $fcc$ and add the total number of valence electrons $NE_{tot}$ based on the chemical intuitions. The final set of 12 primary features that we selected are shown and illustrated in Table~\ref{tab:features}. 

The first few trial trainings, including all compounds, showed no strong correlations, which suggested that the current primary feature set is insufficient to model the whole data set. Based on our chemical insight, we further categorized all compounds into several groups based on the complexity of their structure types. Specifically, we had structure types PbFCl, ZrSiS (ZrSiTe, AmTe$_{2-x}$, and UP$_2$), PrOI and Cu$_2$Sb in group 1, ZrCuSiAs-HfCuSi$_2$, LaZn$_0.5$Sb$_2$ and PrNiSb$_2$ in group 2, CaBe$_2$Ge$_2$ in group 3, and the rest in group 4. Group 1 has only one square net, while Group 2 has two. Group 3 shows different band structures from others, and the remainder is in Group 4. Training including all 861 compounds in group 1 and group 2 gave the current stable results with strong correlations. However, there are 18 more compounds in Group 3 and Group 4, which were originally labeled as "yes". We also include these compounds in examining the descriptors as they were originally categorized as TSMs with high confidence.

\section{SM B: Failure of Principal Component Analysis}\label{app:pca}
\begin{figure}[t!]
    \centering
    \includegraphics[width=0.8\textwidth]{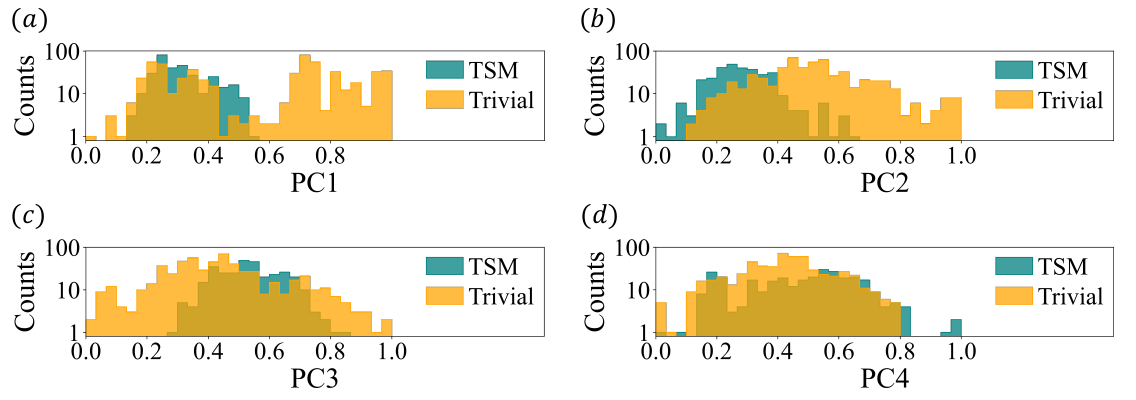}
    \caption{(a)-(d) The distributions of the top 4 principal components (PC's) for the compounds in our data set labeled as TSMs (teal) and trivial (orange) materials. The PC's only reflect the structure of the primary feature set without any information about the labels.}
    \label{fig:PCA}
\end{figure}

In this section, we examine the conventional dimensionality reduction technique, the principle component analysis (PCA). We find that PCA on the primary feature set fails to capture the essential properties that distinguish TSMs from trivial materials.  Fig.\ref{fig:PCA} shows the resulting top 4 principal components. The top 4 principal components fail to separate TSM from trivial materials. This is unsurprising, as the PCA discards all the prior knowledge of observed class labels.

\section{SM C: Method}\label{app:methods}

We apply the Gaussian process (GP) model to discover the correlation between different input dimensions based on the classification performance. However, directly applying GP to classification labels suffers from analytical intractability \cite{Rasmussen2006a}. Thus, we apply transformations to the raw outputs closely following ref. \cite{MiliosNEURIPS2018}, and we describe the details here for completeness.

The labeled data set $\mathcal{D}=\left(\mathbf{X},\mathbf{y}\right)$ for the N materials consists of the input feature matrix $\mathbf{X}=\{\mathbf{x}_1,\dots,\mathbf{x}_N\}$  where  $\mathbf{x}_n\in\mathbb{R}^D$ is the feature vector of the $n^{\text{th}}$ material and $D=12$ is the number of features. The $\mathbf{y}=\bigl(y_1,\dots,y_N\bigr)^{T}$ where $y_n\in \{0,1\}$ are the class labels, with $0$ representing the trivial class and $1$ representing the TSM class. 

Let $f(\mathbf{x})\in\{0,1\}$ be a noisy classifier that predicts the class labels for an input vector $\mathbf{x}$. This prediction is associated with a probability. For the labeled data, the two probabilities are $P\left(f(\mathbf{x}_n)\mid y_n=1\right)$ and $P\left(f(\mathbf{x}_n)\mid y_n=0\right)$. For ease of notation, we denote these two probabilities as:
\begin{eqnarray}
    P_1^{(f_n)} &\equiv& P\left(f(\mathbf{x}_n)\mid y_n=1\right)\\
    P_0^{(f_n)} &\equiv& P\left(f(\mathbf{x}_n)\mid y_n=0\right) 
\end{eqnarray}
where $f_n\equiv f(\mathbf{x}_n) \in \{0,1\}$ are the predicted class labels. Indeed, 
\begin{eqnarray}
    P_i^{(0)}=1-P_i^{(1)}, \text{ for } i=0,1. \label{eq:P_i^0}
\end{eqnarray} 
hence we only discuss $P_i^{(1)}$ in the following steps.

We start with a reasonable model for the distribution of $P_i^{(1)}$ as a Beta function with parameter $\alpha$ given by 
\begin{align}
    P_1^{(1)}\sim \mathrm{Beta}(\alpha+1, \alpha)\label{eq:beta_1}\\
    P_0^{(1)}\sim \mathrm{Beta}(\alpha, \alpha+1)\label{eq:beta_2}
\end{align}
We expect the $P_1^{(1)}$ to be concentrated near one and $P_0^{(1)}$ near 0, as the predicted label $f_n$ should most likely give the assigned label $y_n$; hence, the value of $\alpha$ should be small. As shown in Appendix\ref{app:stability}, we arrive at the optimal value as $\alpha=0.01$ for this work. 

In the next step, we transform this Beta distribution to a Gamma distribution, which is more suited for a Gaussian approximation. Using $\{P_{0}^{(1)},P_{1}^{(1)}\} \in [0,1]$, we construct two random variables $\{\pi_{0}^{(1)},\pi_1^{(1)}\} \in [0,\infty)$, such that
\begin{eqnarray}
P_i^{(1)}=\frac{\pi_i^{(1)}}{\pi_0^{(1)}+\pi_1^{(1)}}  \text{ for } i=0,1.
\end{eqnarray}
 Here, $\{\pi_0^{(1)},\pi_1^{(1)}\}$ are two independent random variables drawn from their respective Gamma distributions given by
\begin{eqnarray}
\pi_i^{(1)} \sim \text{Gamma}\left(\alpha_i^{(1)},1\right)
\end{eqnarray}
where the shape parameter $\alpha_i^{(1)}$ is given by,
\begin{eqnarray}
\alpha_1^{(1)}=1+\alpha,\quad \alpha_0^{(1)}=\alpha.\label{eq:alpha^1}
\end{eqnarray}
We note that with this step, we have generated two independent Gamma distributed random variables. The same treatment applied to $P_i^{(0)}$ (Eq.\eqref{eq:P_i^0}) gives 
\begin{eqnarray}
\pi_i^{(0)} \sim \text{Gamma}\left(\alpha_i^{(0)},1\right) \text{ for } i=0,1,
\end{eqnarray}
where, 
\begin{eqnarray}
\alpha_1^{(0)}=\alpha,\quad \alpha_0^{(0)}=1+\alpha.\label{eq:alpha^0}.
\end{eqnarray}

We are now ready to introduce the  first and only approximation: the Gamma distributed $\pi_i^{(f)}$, where $f\in \{0,1\}$, is approximated with a log-normal distributed $\tilde{\pi}_i^{(f)}$, given by, 
\begin{eqnarray}
\log\left(\tilde{\pi}_i^{(f)}\right)\sim \mathcal{N}\left(\tilde{y}_i^{(f)},\left(\sigma_i^{(f)}\right)^2\right)  \text{ for } i=0,1,
\end{eqnarray}
with $\tilde{y}_i^{(f)}$ and $\sigma_i^{(f)}$ determined from matching the mean and variance of $\Tilde{\pi}_i^{(f)}$ with $\pi_i^{(f)}$. For $\pi_i^{(f)}$, the mean $(\mathbb{E})$ and variance $(\mathbb{V})$ are  given by,
\begin{eqnarray}
\mathbb{E}\left[\pi_i^{(f)}\right]=\mathbb{V}\left[\pi_i^{(f)}\right]=\alpha_i^{(f)},\label{eq:mean_gamma}
\end{eqnarray}
where $\alpha_i^{(f)}$ is given by Eqs.\eqref{eq:alpha^1} and \eqref{eq:alpha^0}. On the other hand, the mean and variance of $\tilde{\pi}_i^{(f)}$ are given by 
\begin{align}
\mathbb{E}\left[\Tilde{\pi}_i^{(f)}\right]&=e^{\tilde{y}_i^{(f)}+\left(\sigma_i^{(f)}\right)^2/2}\label{eq:mean_logN}\\
\mathbb{V}\left[\Tilde{\pi}_i^{(f)}\right]&=\left(e^{\left(\sigma_i^{(f)}\right)^2}-1\right)\left(e^{2\tilde{y}_i^{(f)}+\left(\sigma_i^{(f)}\right)^2}\right)\mathrm{.}\label{eq:var_logN}
\end{align}
Equating Eq.\eqref{eq:mean_gamma} with Eq.\eqref{eq:mean_logN} and \eqref{eq:var_logN}, we get:
\begin{align}
    \Tilde{y}_i^{(f)}&=\mathrm{log}(\alpha_i^{(f)})-\mathrm{log}\left(1/\alpha_i^{(f)}+1\right)/2,\label{eq:tilde_y}\\
\sigma_i^{(f)}&=\left(\mathrm{log}\left(1/\alpha_i^{(f)}+1\right)\right)^{1/2}.\label{eq:sigma_i}
\end{align}
In this work, with our choice of $\alpha=0.01$ as justified in appendix\ref{app:stability}, we obtain $\Tilde{y}_1^{(1)}=\Tilde{y}_0^{(0)}\approx-0.33$, $\left(\sigma_1^{(1)}\right)^2=\left(\sigma_0^{(0)}\right)^2\approx0.69$ and $\Tilde{y}_0^{(1)}=\Tilde{y}_1^{(0)}\approx-4.9$, $\left(\sigma_1^{(0)}\right)^2=\left(\sigma_0^{(1)}\right)^2\approx4.6$. 

From our original class labels $\mathbf{y}=(y_1,\dots,y_N)^{T}$, we can now obtain two transformed vectors $\{\tilde{\mathbf{y}}_0, \tilde{\mathbf{y}}_1\}$ with 
\begin{eqnarray}
\tilde{\mathbf{y}}_i=\left(\tilde{y}_i^1, \dots,\tilde{y}_i^N\right)^{T} \text{ for } i=0,1, 
\end{eqnarray}
whose components $\tilde{y}_i^n\equiv \tilde{y}_i^{(y_n)}$ are given by Eq.\eqref{eq:tilde_y}. The transformed outputs $\tilde{\mathbf{y}}_i$ can be treated as sampled from two independent Gaussian distributed variables. We can use the standard machinery of the Gaussian Process with two latent functions $\tilde{\mathbf{f}}_i=\bigl(\tilde{f}_i(\mathbf{x}_1),\dots,\tilde{f}_i(\mathbf{x}_N)\bigr)^{T}$ to model them. 

With our labeled dataset $\mathcal{D}$, the marginal likelihood for the two GP follows from 
\begin{align}
p\left(\Tilde{\mathbf{y}}_i\mid\mathcal{D},\theta_i\right)&=\int p\left(\Tilde{\mathbf{y}}_i\mid\tilde{\mathbf{f}}_i,\mathcal{D}\right)p\left(\tilde{\mathbf{f}}_i\mid\mathcal{D},\theta_i\right)d\tilde{\mathbf{f}}_i\label{eq:marginal_eq1}
\end{align}
where the model parameters $\theta_i$, the likelihoods $p\left(\Tilde{\mathbf{y}}_i\mid\tilde{\mathbf{f}}_i,\mathcal{D}\right)$ and the priors $p\left(\tilde{\mathbf{f}}_i\mid\mathcal{D},\theta_i\right)$ are described below. The likelihoods in the two GP's are Gaussians, 
\begin{align}
p\left(\Tilde{\mathbf{y}}_i\mid\tilde{\mathbf{f}}_i,\mathcal{D}\right)=\mathcal{N}(\Tilde{\mathbf{y}}_i\mid\tilde{\mathbf{f}}_i, \mathbf{\Sigma}_i) \text{ for } i=0,1,\label{eq:likelihood}
\end{align} 
where $\mathbf{\Sigma}_i$ are  diagonal matrices with elements $(\mathbf{\Sigma}_i)_{n,n}=\left(\sigma_i^{(y_n)}\right)^2$ given by Eq.\eqref{eq:sigma_i}. The priors of the latent functions are also Gaussians, 
\begin{eqnarray}
p\left(\tilde{\mathbf{f}}_i\mid\mathcal{D},\theta_i\right)=\mathcal{N}(\tilde{\mathbf{f}}_i\mid\bm{\mu}_i, \mathbf{K}),  \text{ for } i=0,1,\label{eq:prior}
\end{eqnarray} 
where $\bm{\mu}_i=\mu_i(1,\dots,1)$ is a parameter and $\mathbf{K}$ is the covariance matrix (same for both priors) whose elements are given by the kernel function $\mathbf{K}_{mn}=k(\mathbf{x}_m, \mathbf{x}_n)$. The kernel function for two feature vectors $\mathbf{x}$ and $\mathbf{x}'$ is defined as
 \begin{equation}
     k(\mathbf{x}, \mathbf{x}'):=\sigma_p^2 \exp\Bigl[-\frac{1}{2}(\mathbf{x}-\mathbf{x}')^T\bigl({\mathbf{L}^T\mathbf{L}+\mathbf{\Lambda}}\bigr)(\mathbf{x}-\mathbf{x}')\Bigr]\mathrm{,}\label{eq:kernel}
 \end{equation}
 where $\mathbf{L}$ is a $q\times D$  matrix ($q=6$ in this work, see appendix~\ref{app:stability}), and $\mathbf{\Lambda}$ is a diagonal matrix with only non-negative elements. Thus the set of all hyperparameters for the two GP is $\theta_i=\{\bm{\mu}_i, \mathbf{L}, \mathbf{\Lambda}, \sigma_p\}$. They are referred to as hyperparameters because GP is a non-parametric model which already integrates out the usual parameters of the latent function.  
 
 Thanks to the Gaussian likelihood and prior, the two marginal likelihoods in Eq.\eqref{eq:marginal_eq1} have a closed form expression given by,
 \begin{align}
p\left(\Tilde{\mathbf{y}}_i\mid\mathcal{D},\theta_i\right)=\mathcal{N}\left(\Tilde{\mathbf{y}}_i\mid\bm{\mu}_i,\mathbf{\Sigma}_i+\mathbf{K}\right) \text { for } i=0,1. &
 \end{align}
Now we proceed as in any GP and learn the relevant features in data by maximizing the log marginal likelihood (LML)\cite{Rasmussen2006a}. We start our training by initializing $\mathbf{L}$ with its elements drawn from a standard Gaussian distribution and divided by $q$ (the reduced dimension of $\mathbf{L}$ in Eq.\eqref{eq:kernel}), given by $\mathbf{L}_{ij}\sim\mathcal{N}(0,1)/q$. At the same time, all diagonal entries in $\mathbf{\Lambda}$ are initialized to $1$. With our model structure set by the hyperparameters $\alpha$ (Eq.~\eqref{eq:beta_1},\eqref{eq:beta_2}) and $q$, the LML  with parameters $\theta\equiv\{\theta_1,\theta_2\}$ is given by,
\begin{flalign}
    \text{(LML)}_{\alpha,q}(\theta)&=\sum_{i=0}^1\mathrm{log}(p(\Tilde{\mathbf{y}}_i\mid\mathcal{D},\theta_i))\nonumber\\
    &=-\frac{1}{2}\sum_{i=0}^1\;\Bigl[(\Tilde{\mathbf{y}}_i-\bm{\mu}_i)^T\bigl(\mathbf{K}+\mathbf{\Sigma}_i\bigr)^{-1}(\Tilde{\mathbf{y}}_i-\bm{\mu}_i)+\mathrm{log}|\mathbf{K}+\mathbf{\Sigma}_i|+N\mathrm{log}2\pi\Bigr],&
\end{flalign}
where $|(\dots)|$ denotes the matrix determinant. Minimizing negative LML through gradient descent gives the optimal parameters $\theta^*_{\alpha,q}$, 
\begin{eqnarray}
    \theta^*_{\alpha,q} = \text{argmin}_{\theta}\left(\text{(-LML)}_{\alpha,q}(\theta)\right)
\end{eqnarray}
From the learned parameters in $\theta^*_{\alpha,q}$, the quantity of particular interest for us is the covariance matrix of the features given by,
\begin{eqnarray}
\mathbf{M}^*=\mathbf{L}^{T}\mathbf{L}+\mathbf{\Lambda}.\label{SM_Eq:M*}
\end{eqnarray} 

To reflect how PFs interact to form the descriptors of TSM, we construct a normalized correlation with elements given by
\begin{eqnarray}
\mathbf{C}_{mn}=\dfrac{\mathbf{M}^*_{mn}}{\sqrt{\mathbf{M}^*_{mm}\mathbf{M}^*_{nn}}}.\label{SM_eq_norm_corr}
\end{eqnarray}
\makeatletter 
\makeatother

\section{SM D: Hyperparameter and Stability Testing}\label{app:stability}
\begin{figure*}[ht!]
    \centering
    \includegraphics[width=0.95\textwidth]{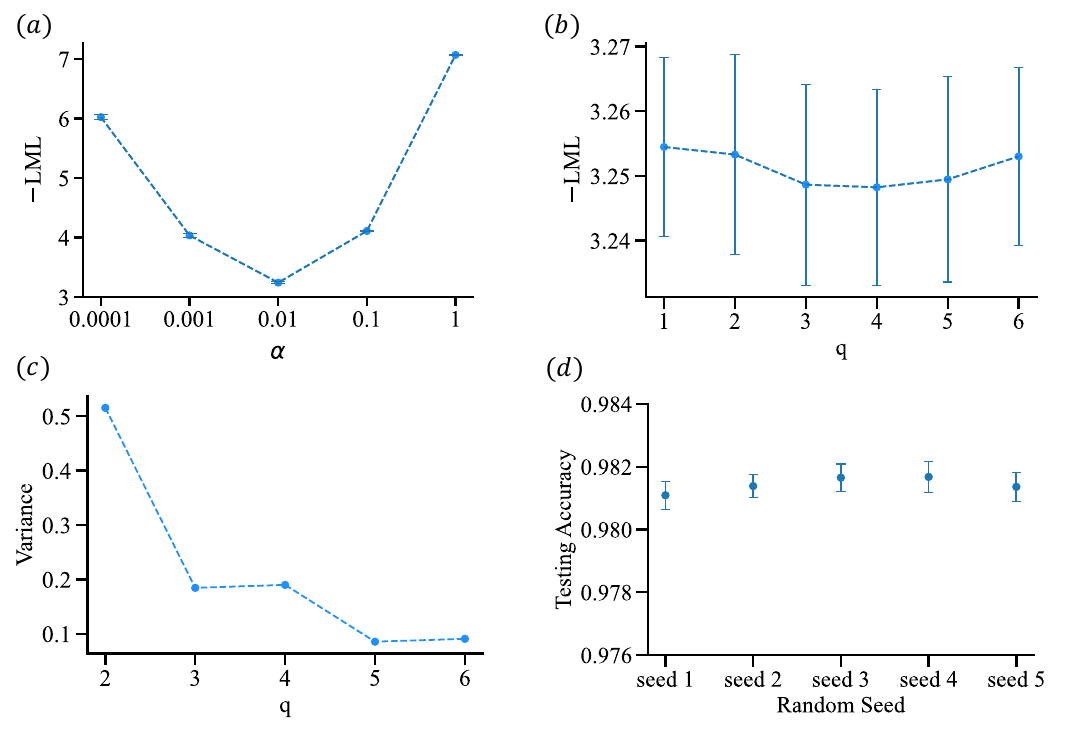}
    \caption{(a) The $-$LML values for different $\alpha$'s when $q=6$. The minimum at $\alpha=0.01$ is consistent for all $q$'s. The symbols show the mean, and the error bars show one standard deviation over 60 trials. (b) The $-$LML values for different $q$'s when $\alpha=0.01$. The symbols show the mean, and the error bars show one standard deviation over 60 trials. The optimal choice of $q$ is not obvious from here. (c) The sum of the variances of all elements in the $\mathbf{C}$ matrix divided by 2 when $\alpha=0.01$, for different $q$'s. This plot shows that $q=6$ is a safe choice for a stable $\mathbf{C}$ matrix. (d) The classification accuracies are given by the trained model when $\alpha=0.01$ and $q=6$, for 5 experiments with different random seeds. The symbols show the mean, and the error bars show one standard deviation over 60 trials.}
    \label{fig:Hyper}
\end{figure*}

Every time we train the model, we apply 5-fold cross-validation (CV) twelve times and average the results. We determine the hyperparameters $q$ (for the $L$ matrix) and $\alpha$, by inspecting the negative log marginal likelihood (-LML) performances on the test sets with different $q$ ($\alpha$ fixed) and $\alpha$ ($q$ fixed), under the constraints that $q\leq6$ and $\alpha\leq1$. The constraint for $q$ controls the number of hyperparameters in a reasonable range, and the constraint for $\alpha$ keeps the Beta distributions fairly sharp. To proceed, we iteratively fixed one and varied the other until we determined the optimal choice to be $q=6$ and $\alpha=0.01$. In Fig.\ref{fig:Hyper} (a), we show that the $-$LML is minimized by the choice $\alpha=0.01$, which is consistent for all values of $q$. Having $\alpha=0.01$, we obtained the flat behavior of the $-$LML for all $q$'s, as shown in Fig\ref{fig:Hyper} (b). We further explored the variance of the correlation matrix $\mathbf{C}$ (Eq.~\eqref{SM_eq_norm_corr}), with different $q$'s. Fig\ref{fig:Hyper} (c) shows the sum of variances for all elements in the correlation matrices given by five experiments with different random seeds for different $q$'s. Increasing $q$ until $q=6$ has an obvious effect on reducing the fluctuations in the correlation matrices. We also demonstrate the stability of our model in Fig. \ref{fig:Hyper} (d). The classification accuracies fluctuate between $98.0\%$ and $98.2\%$ for 5 experiments using different random seeds. With the results in (c) and (d), we conclude that our model learns stable and robust criteria for identifying square-net TSMs. The codes used in this paper are based on the Python package GPyTorch \cite{GardnerNEURIPS2018}. 

We point out that in reference \cite{MiliosNEURIPS2018}, the author suggests that the marginal likelihood is not a good metric for determining the optimal $\alpha$ when the GP applies a single zero-mean prior. We avoid this problem by introducing two different constant-mean priors. Then, we obtain $\alpha=0.01$ from the log marginal likelihood values.

\newpage
\section{SM E: Strong Correlations}\label{app:correlations}

\begin{figure}[t!]
    \centering
    \includegraphics[width=0.9\textwidth]{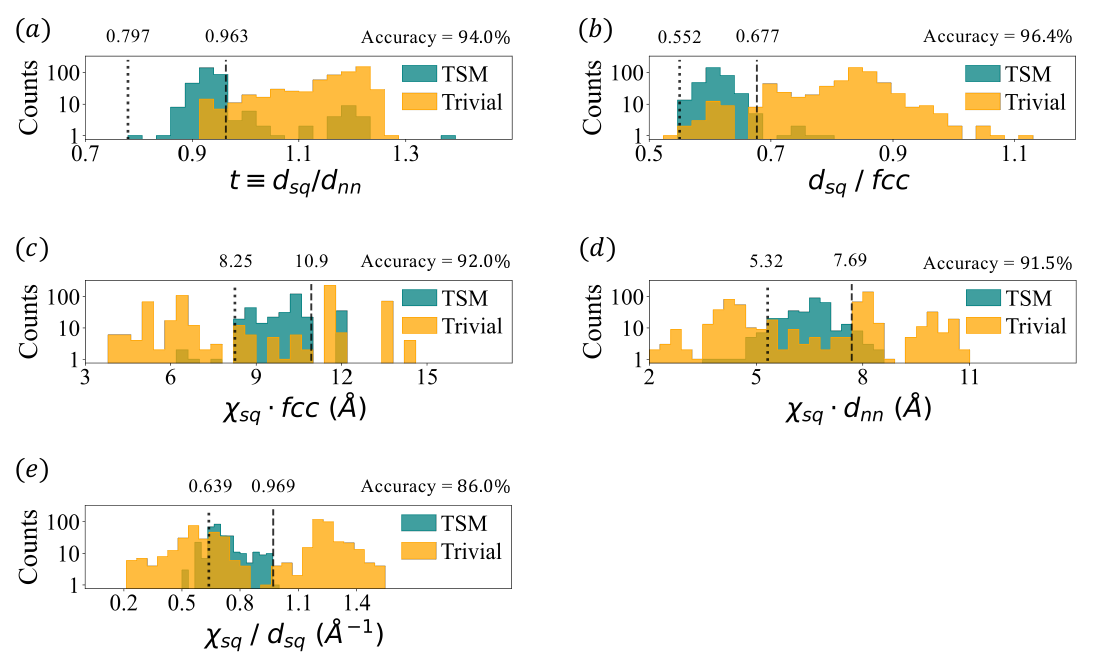}
    \caption{The distributions of (a): $d_{sq}/d_{nn}$, (b): $d_{sq}/fcc$, (c): $\chi_{sq}\cdot fcc$, (d): $\chi_{sq}\cdot d_{sq}$, (e): $\chi_{sq}/d_{sq}$ for the TSM and trivial materials in our dataset. The vertical lines define an interval for each plot such that classifying all materials inside the interval as TSMs, and all materials outside the interval as trivial, would maximize the classification accuracy. The boundaries of the interval and the accuracy according to such criteria for each panel are given.}
    \label{fig:1d}
\end{figure}

Among all the off-diagonal elements in the normalized-correlation matrix $\boldsymbol{C}$ (Fig. 3(a) of main text), there are five explicitly strong correlations, which are ($d_{sq}$, $d_{nn}$), ($d_{sq}$, $fcc$), ($\chi_{sq}$, $d_{sq}$), ($\chi_{sq}$, $d_{nn}$), and ($\chi_{sq}$, $fcc$). We take either their products or ratios based on their signs in $\boldsymbol{C}$ and obtain an emergent descriptor for each of these pairs.

We then examine the 1D classification results using these candidates, as shown in Fig.~\ref{fig:1d}. We quantify their classification power with their ``predictive accuracy'' which we define as the maximum accuracy attained by classifying the materials inside a closed interval as TSMs. The maximum accuracy is obtained by optimizing the lower and upper bounds of the interval. The values of the upper and lower bounds separating TSMs from trivial materials and their accuracy in separation are shown in Fig.~\ref{fig:1d}.


This classification strategy introduces some bias into the results because the total numbers of TSMs and trivial materials are not balanced, and their distributions in each candidate's space are also different. However, it still qualitatively shows that emergent descriptors have obvious classification power. 

Of the five descriptors, $\chi_{sq}/d_{sq}$   carries similar information as $\chi_{sq}\cdot fcc$. Both descriptors predict the likeliness of hypervalent-bond formation between square net elements. While $\chi_{sq}/d_{sq}$ caries the information about the relationship between actual bond lengths found within the data set, $\chi_{sq}\cdot fcc$ carries more general information, relating atom's size to its electronegativity, to predict the probability of formation of hypervalent bonds. Additionally, $\chi_{sq}\cdot fcc$ contains only atom-specific information, affording greater predictive power. The similarities between $\chi_{sq}\cdot fcc$ and $\chi_{sq}/d_{sq}$ are also indicated in their 1D classification plots (Fig~\ref{fig:1d} (c) and (e)). In both plots, most of the TSMs group in a single cluster somewhere in the middle. Thus, based on the similarities in the chemical meaning and data distribution, we only keep $\chi_{sq}\cdot fcc$ in our later analysis.

\section{SM F: Hypervalency signatures of TSM from ($\chi_{sq},d_{nn}$) and ($\chi_{sq},fcc$) }

We gain a consistent picture by going back to the pairs of primary features $(\chi_{sq},d_{nn})$ and $(\chi_{sq},fcc)$ that generated the emergent descriptors $\chi_{sq}*d_{nn}$ and $\chi_{sq}*fcc$. We utilize our GP model to obtain the probability of getting TSMs in the 2D spaces spanned by $(\chi_{sq},d_{nn})$ as well as by $(\chi_{sq},fcc)$. 

We can use our GP model to get probabilities after we learn the hyperparameters of the GP. The posterior distribution of the latent function $\Tilde{f}^*_i\equiv\Tilde{f}_i(\mathbf{x}^*)$ at a new input point $\mathbf{x}^*$ is
\begin{equation}
    p(\Tilde{f}^*_i\mid\mathcal{D},\theta_i)=\int p(\Tilde{f}^*_i\mid\Tilde{\mathbf{f}}_i)p(\Tilde{\mathbf{f}}_i\mid\mathcal{D},\theta_i)d\Tilde{\mathbf{f}}_i
\end{equation}
where $p(\Tilde{f}^*_i\mid\Tilde{\mathbf{f}}_i)$ is the joint Gaussian distribution given by the prior just as $p(\Tilde{\mathbf{f}}_i\mid\mathcal{D},\theta_i)$. We can then estimate the predictive  probabilities of the class labels at $\mathbf{x}^*$ by integrating over the latent functions:
\begin{equation}
    \mathbb{E}\left[P^*(\mathrm{TSM})\right]=\int \frac{\mathrm{exp}(\Tilde{f}^*_1)}{\sum_{j=0,1}\mathrm{exp}(\Tilde{f}^*_j)}p(\Tilde{f}^*_1\mid \mathcal{D},\theta_i)d\Tilde{f}^*_0d\Tilde{f}^*_1
\end{equation}
where $P^*(\mathrm{TSM})=P(\mathrm{TSM}\mid\mathbf{x}^*)$. 

To get the probability of TSM ($P_{TSM}$) in the 2D spaces spanned by $(\chi_{sq},d_{nn})$, as well by  $(\chi_{sq},fcc)$, we trained the GP model with only 4 PF's, $\chi_{sq}$, $d_{sq}$, $d_{nn}$, and $fcc$, which effectively projects our input data to the four-dimensional space spanned by these PF's. Then, for $P_{TSM}$ of two PFs, we fixed the remaining two PFs to their mean values and plotted the predictive probabilities of TSMs given by Eq. (31) for different values of the two PFs being studied. We eliminated the rest 8 PFs in this process because their unbiased mean values are ill-defined. This is because the distributions of both TSMs and trivial materials are not uniform, and bothering with the fixed values of the rest 8 PFs might introduce a strong bias in the prediction.

\begin{figure}[t!]
    \centering
    \includegraphics[width=0.9\textwidth]{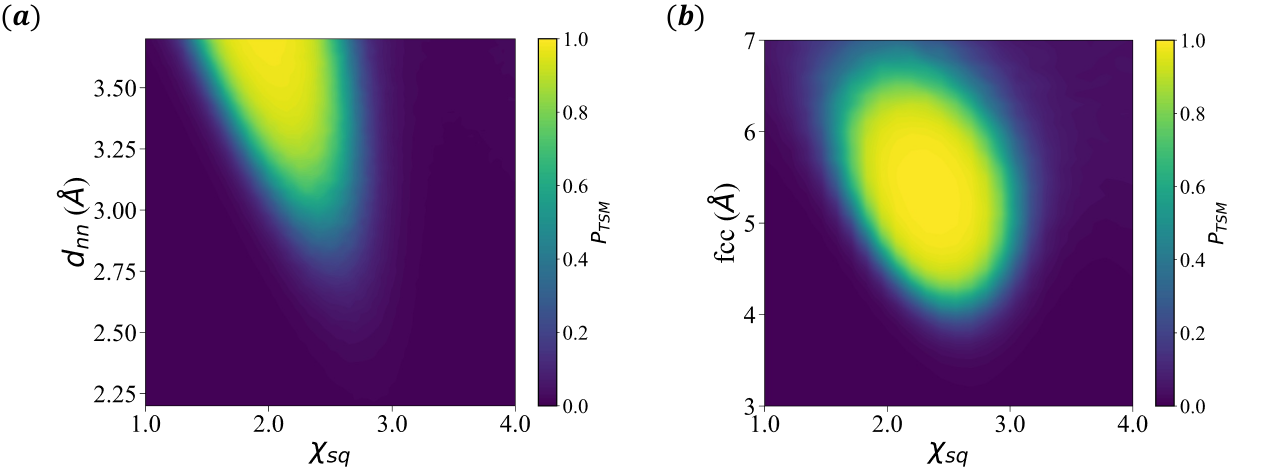}
    \caption{The model's prediction of the probability $P_{\rm{TSM}}$ for square-net materials being TSMs in the 2D subspace spanned by ($\chi_{sq}, d_{nn}$) in panel  {\bf{(a)}}, and in ($\chi_{sq}, fcc$) space in panel  {\bf{(b)}} . }
    \label{fig:2d_prob}
\end{figure}

The high probability (bright yellow) region of $P_{\rm{TSM}}(\chi_{sq},d_{nn})$ in Fig~\ref{fig:2d_prob}(a) clearly shows that large values of $d_{nn}$ and intrmediate values of $\chi_{sq}$ favor TSM. This agrees well with our intuition for TSM materials: small values of $d_{nn}$ hinders the formation of TSM by favoring the square-net atoms to form bonds with out-of-plane atoms, while large $d_{nn}$ with intermediate electronegativity facilitates hypervalent square-net bonds enabling TSM. 

\para The high probability region in $P_{\rm{TSM}}(\chi_{sq},fcc)$ (Fig~\ref{fig:2d_prob}(b)) also clearly shows that the ideal atoms have intermediate electronegativity and size, in agreement with our chemical heuristics presented in the main text: only elements of intermediate electronegativity and size are inclined to form hypervalent bonds.

\section{SM G: Outlier Compounds}

The separation of compounds into TSM and trivial domains in the 2D space of ($\chi_{sq}\cdot fcc$, $t$) (Fig 3(h) of main text) identified some materials that were mislabeled as topological or trivial in the original data set. We have identified 44 of these outlier compounds. We ran calculations on 29 pure compounds. Since the band structures of alloys are not reported in TMBb and were inferred using the band structure of a pure compound with a similar chemical composition in the original data set, we have repeated the same procedure for the remaining 15 compounds.

Calculations on the originally incorrectly labeled compounds were done using the Vienna Ab initio Simulation Package (VASP) 5.4.4. software \cite{Kresse1996CMS, Kresse1996PhysRevB} following the procedure used to calculate bands in the Topological Materials Database.\cite{Vergniory2019Natureb} The electronic structures were calculated using the Perdew-Burke-Ernzerhof (PBE) functional and recommended Projector Augmented Wave (PAW) potentials for all atoms.\cite{Perdew1996PhysRevLett, Blochl1994PhysRevB, Kresse1999PhysRevB} Spin-orbit coupling was included in the calculations.\cite{Steiner2016PhysRevB} The size of a $\Gamma$-centered Monkhorst-Pack mesh \cite{Monkhorst1976PhysRevB} was $11\times11\times11$. We further examined these compounds by projecting the contributions of the square-net atoms' on their band structures using the pymatgen.electronic\_structure package.\cite{Ong2013CMS}

\newpage

\end{document}